\documentclass[prl,twocolumn,final]{revtex4}

\usepackage{graphicx}

\begin{document}

\title{Response of Cloud Condensation Nuclei ($>$ 50 nm) to changes in ion-nucleation}

\author{Henrik Svensmark}
\email[hsv@space.dtu.dk]{}
\author{Martin B. Enghoff}
\email[enghoff@space.dtu.dk]{}
\author{Jens Olaf Pepke Pedersen}
\email[jopp@space.dtu.dk]{}

\thanks{}
\affiliation{Center for Sun-Climate Research, National Space Institute, Technical University of Denmark, Juliane Maries Vej 30, 2100 Copenhagen {\O}, Denmark}
\received{December 2011}

\begin{abstract}
In experiments where ultraviolet light produces aerosols from trace amounts of ozone, sulphur dioxide, and water vapour, the number of additional small particles produced by ionization by gamma sources all grow up to diameters larger than 50 nm, appropriate for cloud condensation nuclei. This result contradicts both ion-free control experiments and also theoretical models that predict a decline in the response of larger particles due to an insufficiency of condensable gases (which leads to slower growth) and to larger losses by coagulation between the particles. This unpredicted experimental finding points to a process not included in current theoretical models, possibly an ion-induced formation of sulphuric acid in small clusters.
\end{abstract}
\pacs{92.20.Bk,92.70.Gt,96.40.-Z}

\maketitle
The role of ionization in atmospheric processes has been a controversial matter since it was suggested fifty years ago\cite{ne59,Dickinson1975}, and found in the correlation between global cloud cover and the influx of galactic cosmic rays (GCR)\cite{Svensmark1997JATP}. Subsequent studies have shown correlations between GCR variations and changes in aerosol counts and cloud properties in the atmosphere\cite{ha05,Svensmark2009GeoRL,JSvensmark2012}, but these are still disputed\cite{Kristjansson2008,Sloan2008,Calogovic2010}.

Fortunately, the issue can also be addressed in the laboratory. Experimental evidence for a microphysical mechanism was first reported in 2007\cite{Svensmark2007RSPSA} and further experiments have recently added to its credibility\cite{Enghoff2011GRL,cloud2011}. These experiments initially showed that an increase in ionization leads to an increase in the formation of ultra-fine aerosols ($\approx$ 3 nm), but in the real atmosphere such small particles have to grow by coagulation and intake of condensable gases to become cloud condensation nuclei (CCN) ($>$ 50 nm) in order to have an effect on clouds\cite[chapter 17]{seinfeld2006}.

Theoretical doubts about the likelihood of such particle growth into CCN have arisen from consideration of (1) the competition between the additional ultra-fine aerosols for the limited supply of condensable gases leading to a slower growth and (2) the larger losses of the additional particles during the longer growth-time to larger particles by coagulation and by other loss mechanisms. Indeed numerical studies using the current knowledge of aerosol dynamics predict that variations in the count of ultra-fine aerosols will lead only to an insignificant change in the count of CCN\cite{Pierce2009,Snow-Kropla2011}. It is even suggested that an increased production of ultra-fine particles as a result of GCR ionization leads to a reduction in the CCN count.

\begin{figure}
\includegraphics[scale=1.]{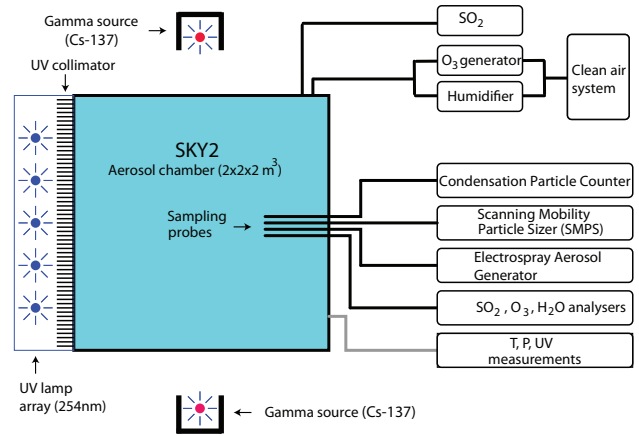}
\caption{Schematic diagram of the SKY2 experiment.}
\label{SKYfig}
\end{figure}

In order to study the growth of aerosols to CCN sizes, measurements were performed in an 8 m$^3$ reaction chamber (SKY2) made from electro-polished stainless steel shown schematically in Fig.\ref{SKYfig}. One side was fitted with a Teflon foil to allow ultraviolet light (253.7 nm) to illuminate the chamber, which was continuously flushed with dry purified air. Variable concentrations of water vapor (H$_2$O), ozone (O$_3$), and sulphur dioxide (SO$_2$) could be added to the chamber, where the pressure was held a few Pa above atmospheric pressure, and the temperature at around 296 K. The UV-lamps initiated a photochemical reaction producing sulphuric acid (H$_2$SO$_4$).

Ions were produced in the chamber by the naturally occurring GCR and by background radiation from radon, and the ionization could be enhanced with two Cs-137 gamma sources (30 MBq), mounted on each side of the chamber. The total number of aerosols generated in the chamber were measured with a TSI Model 3775 Condensation Particle Counter (CPC) with a cutoff at 4 nm. A particle size measurement was done with an electrostatic classifier (TSI model 3080) fitted with a nano-DMA (TSI model 3085) covering the range 3-65 nm and a CPC (TSI 3025A). Ultra-fine H$_2$SO$_4$-water aerosols ($\approx$ 10 nm) could be generated with an Electrospray Aerosol Generator (TSI model 3480).

Concentrations of ozone were measured with Teledyne T400 analyzer and sulfur dioxide with a Thermo 43 CTL analyzer. The chamber was also equipped with instruments to measure temperature, differential and absolute pressure, humidity, and UV intensity.


Estimates of the sulphuric acid concentrations were made by measuring the growth-rate of particle diameters above 3 nm\cite{Kulmala2001}.
This method gave typical concentrations in the range 1-10  ppt, i.e., $\sim$ 10$^7$ - 10$^8$ molecules cm$^{-3}$.

The experiments were run in a mode where steady state conditions of H$_2$SO$_4$ were achieved under continuous exposure to the UV-light. Typically the gas mixture in the reaction chamber consisted of 40-50 ppb O$_3$, 0.8-1.0 ppb SO$_2$ and a relative humidity of 25\%.
\begin{figure}
\includegraphics[angle=90,scale=.4]{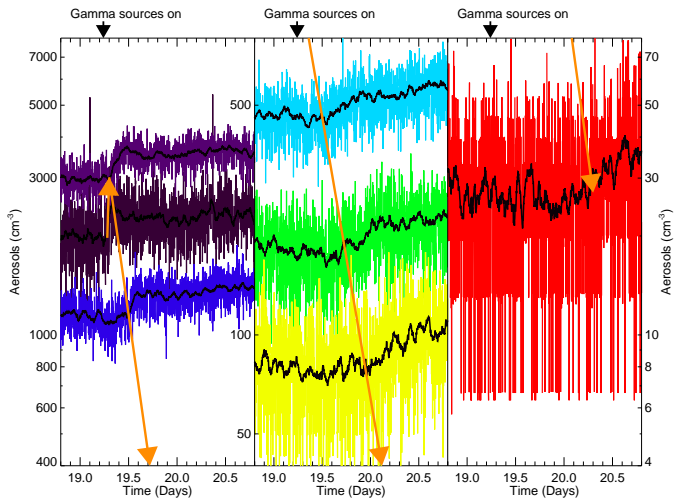}
\caption{During a typical experimental run, number densities of aerosol particles of increasing diameter were observed as a function of time. Left panel:- 3-10 nm (black), 10-20 nm (purple), 20-30 nm (dark blue). Middle panel: 30-40 nm (light blue), 40-50 nm (green), 50-60 nm (yellow). Right panel: 60-68 nm (red). At $\approx$ 19.2 days the gamma sources were opened to increase the ionization as described in the text, and an increase in aerosol density began immediately in the 3-10 nm curve (black). Subsequently the increase in number densities slowly propagated down to the larger aerosol sizes, as shown by the slanting arrow. Note that the number of particles in the first bin (3-10 nm) is relatively low because of a lower sensitivity of the instrument to the smallest aerosols. Black curves are an average over 67.5 minutes.}
\label{Exptypicalrun}
\end{figure}

Experiments  were  performed  under various levels of ionization and UV intensity in the chamber. After changing one of these parameters the aerosols were allowed to grow and conditions to settle to a new steady state for a period of 24-36 hours. For example the gamma sources were opened resulting in an increase in ionization from about 3 ions-pairs cm$^{-3}$s$^{-1}$ to 60 ions-pairs cm$^{-3}$s$^{-1}$. This resulted in an increase of about 20 \% in the formation of small aerosols. The parameters were then kept constant for a period of about 36 hours until the new steady state was achieved.

Figure \ref{Exptypicalrun} shows such a run, where the number densities of seven sizes of aerosols, from 6.5 nm to 64 nm, are plotted as a function of time. When the gamma sources were opened, at $\approx$19.2 days, an increase in aerosol density was seen to follow directly in the 3-10 nm curve (black) and then to propagate slowly through increasing aerosol diameters down to the larger aerosol sizes. After a new steady state was reached, the gamma sources were closed, and a small decrease in aerosol density could be monitored, again starting quickly with the small aerosols and propagating gradually to the larger aerosols (not shown in Fig. \ref{Exptypicalrun}).

\begin{figure}
\includegraphics[scale=.6]{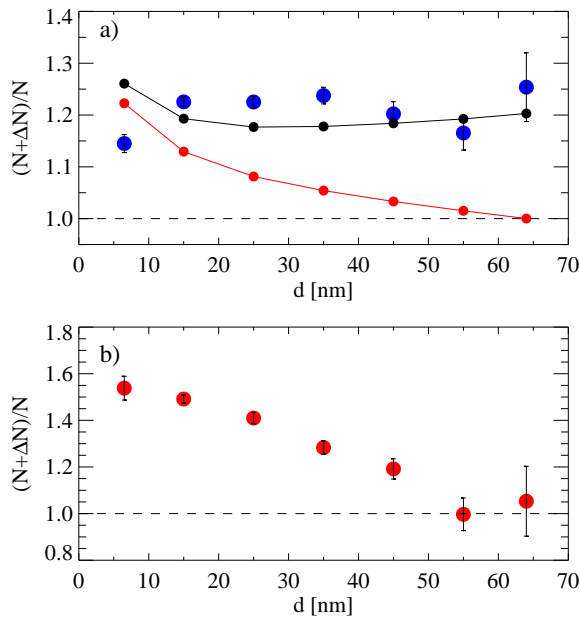}
\caption{Steady state response to a change in nucleation as a function of particle diameter, normalized to the particle number before two types of perturbation. a) Ion-induced increase in nucleation. Blue circles are the experimental results averaged over five runs. The red curve is a typical result of a numerical simulation of the experimental situation using a standard numerical aerosol model. Notice that the expected response from the modeling decreases strongly with particle diameter in contrast with the experimental results. A much better agreement is seen with a numerical simulation in the black curve, where the concentration of sulphuric acid is held constant. b) Control experiment where the increase in particle concentration is done by injection of H$_2$SO$_4$-water ultra fine aerosols ($\approx$ 6-8 nm) under constant UV intensity and trace gas concentrations, and no gamma source ionization. Notice that in this case the response (red circles) diminishes as aerosol size increases. Error bars are $\pm$1-$\sigma$ errors.}
\label{Exp_dn}
\end{figure}

For each experimental run the density of particles before (and after) an imposed ionization change was averaged over a period of 2.25 hours (prior to and after the change) and the mean and standard error of the mean was calculated. And finally the change in the response was averaged over five runs. The blue circles in Fig. \ref{Exp_dn}a show the relative response to changes in ion-nucleation as a function of particle size, averaged over the five runs. It is seen that the response is remarkably constant over the shown size range.

It is of interest to contrast the above experiment to a situation without ionization and a \textit{constant} H$_2$SO$_4$ production subject to an increase in ultra-fine aerosols $\approx$ 6-8 nm. The experimental procedure is first to reach steady state conditions using a constant UV intensity and trace gas concentrations as before, followed by a constant injection of H$_2$SO$_4$-water ultra-fine aerosols produced by the Electrospray Aerosol Generator. Figure \ref{Exp_dn}b displays the response, averaged over 3.5 hours, as a function of aerosol sizes to the aerosol injection. It is seen that in this case the response is diminishing as a function of size in accordance with the theoretical expectations.

The experimental results can be compared with numerical simulations of a general dynamics equation of aerosols. The evolution of the cluster distribution is given by\cite{Seinfeld1998}
\begin{eqnarray}\label{aeq}
\frac{\partial N_k}{\partial t}&=& \frac{1}{2}  \sum_{j=2}^{k-1} K_{j,k-j} N_j
N_{k-j}
- \sum_{j=1}^{\infty}  K_{k,j}  N_k  N_j  \nonumber\\
&& - \frac{\lambda}{{r_i}^\gamma}  N_k + \beta_{k-1} N_{k-1} - \beta_{k} N_{k} + S \delta_{k,k_0}
\end{eqnarray}
where $N_k$ is the number density of clusters each containing $k$ sulphuric acid molecules, assuming that the equilibrium concentration of water molecules in each cluster is reached instantaneously\cite{Yu2005}. $K_{k,j}$ is the coagulation coefficient determined from \citet{Laakso2002}, and can be used for all Knudsen numbers and hence from diameters of $<$ 1 nm to $>1$ microns. The
radius $r_i$ of a cluster with $i$ H$_2$SO$_4$ molecules and a number of water molecules depends on the humidity\cite[Chap.10]{seinfeld2006}.  Particle loss to the chamber walls is approximated with the $\lambda/r_i^\gamma$ term where $\gamma$ is determined experimentally to $\gamma =$0.69 $\pm$ 0.05 from
the decay of particles in the chamber, and $\lambda =$  6.2 $\pm$ 2.0 $\cdot$10$^{-4}$  nm$^{\gamma-1}$s$^{-1}$. $S$ is the production of new critical clusters with size given by $k_0$ H$_2$SO$_4$ molecules. The nucleation rate $S$ is either a constant $S=S_0$ or function of the H$_2$SO$_4$ concentration, e.g. $S =$ $\alpha$ [H$_2$SO$_4$]$^2$, where $\alpha$ is a constant. The $\beta_k$-term describes the condensation of H$_2$SO$_4$ molecules in the gas phase to the $k$'th cluster and are found according to \citet{Laakso2002}, with the value 1 of the mass accommodation coefficient\cite{Laaksonen2005}, and a mean free path from \citet{Lehtinen2003}.

The equation governing the sulphuric acid concentration is
\begin{equation} \label{eq10}
\frac{d [H_2SO_4]}{dt} = P_{H_2SO_4} - ( L + \lambda_{H_2SO_4} )[H_2SO_4]
\end{equation}
where $P_{H_2SO_4}$ is the production of gaseous sulphuric acid. The second term $L$ is the
loss of H$_2$SO$_4$ molecules to the aerosols by condensation. The last term is the loss of H$_2$SO$_4$ molecules to the chamber walls, and is determined from extrapolating the size dependent aerosol losses to the size of a H$_2$SO$_4$ molecule to $\lambda_{H_2SO_4} =$ 7.2 $\pm$ 3.0 $\cdot$10$^{-4}$  s$^{-1}$. The model is described in more detail in \citet{Bondo2010}.

The red curve in Fig. \ref{Exp_dn}a shows the result of a numerical simulation of an increase in the nucleation similar to the experimental situation, where the production term of H$_2$SO$_4$ is kept constant, but the concentration of H$_2$SO$_4$ can vary. The response goes slowly to zero with increasing size of the clusters, due to a smaller concentration of H$_2$SO$_4$. This simulation of the aerosol dynamics is consistent with a number of recent simulations which show very small responses at CCN sizes to a change in the nucleation rate. In their MODGIL simulation Pierce and Adams found responses of 0.004\% in CCN (at 0.2\% supersaturation) to a 4-fold increase in new particle formation,  and in another simulation (IONLIMIT) they found a 0.08\% change in CCN (again at 0.2\% supersaturation) to a 24\% increase in nucleation\cite{Pierce2009}.

As the expectation shown in the red curve in Fig. \ref{Exp_dn}a is contradicted by the experimental results (blue circles) an obvious question is whether the ionization by gamma rays may produce sufficient H$_2$SO$_4$ in the gas phase to replace the expected loss of $\approx$ 3-7\% from the additional particles (estimated from numerical simulations). Each ion-pair will on average\cite{Willis1976,Muller1993} produce two OH molecules. Therefore with an ionization of 60 ion-pair cm$^{-3}$s$^{-1}$ the production will be 120 molecules cm$^{-3}$s$^{-1}$. From the experimentally estimated losses and growth rates the production of H$_2$SO$_4$ from the photolysis is 3.5$\cdot$10$^4$ molecules cm$^{-3}$s$^{-1}$. If every OH molecule becomes an H$_2$SO$_4$ molecule its production will be 0.3\% of the photolysis, i.e 10 times lower than estimated loss of H$_2$SO$_4$. From the experiment it is known that only a minor fraction of the OH are consumed in the pathway that leads to H$_2$SO$_4$. It is therefore safe to conclude that the production of H$_2$SO$_4$ by this path is at least an order of magnitude too small to explain the observed. Another main chemical species produced in moist air (N$_2$-O$_2$-H$_2$O) is nitric acid (HNO$_3$) that potentially could help the condensational growth of the aerosols. Experimentally it is found that each ion-pair produces 0.4 HNO$_3$ molecules\cite{Jones1959}, which in the chamber leads to a production of 24 cm$^{-3}$s$^{-1}$ HNO$_3$ molecules. This production is only 0.07\% of the photolysis production of H$_2$SO$_4$, and therefore nearly 40 times to small too explain the experimental results.

In a second simulation, illustrated in the black curve in Fig. \ref{Exp_dn}a, the concentration of H$_2$SO$_4$ is artificially held constant. In this case the response of larger particles to the additional nucleated particles is not going to zero and the match to the experimental results is much better. How, then, is the growth of the particles sustained? The indication from the second numerical simulation is that effectively there is no decrease in the concentration of condensable gases, even though the UV photolysis of H$_2$SO$_4$ is held constant throughout the duration of the experiment. But the additional ion-nucleated particles should effectively decrease the H$_2$SO$_4$ concentration with $\approx$ 3-7 \%.

A possible explanation could be that the charged clusters are producing additional H$_2$SO$_4$ molecules from reactions involving negative ion chemistry of O$_3$, SO$_2$ and H$_2$O, where a negative ion can be reused in a catalytic production of several H$_2$SO$_4$. Such reactions were first suggested in \citet{Svensmark2007RSPSA}, and also in a recent experiment\cite{Enghoff2012} looking at isotope fractionation of sulphur from either UV or from ion-induced generation of H$_2$SO$_4$, where the sulphur isotope fractionation was used to distinguish the different pathways leading to H$_2$SO$_4$. It was found in the presence of ionization alone, that for each ion-pair 27.8$\cdot10^6$ H$_2$SO$_4$ molecules were produced, using extreme gas mixing ratios, i.e. 0.01 \% SO$_2$, 400 pbb O$_3$, 40 \% RH H$_2$O and 1000 ions-pairs cm$^{-3}$s$^{-1}$. Scaling to the present experiment gives a production of 38.9 H$_2$SO$_4$ molecules pr. ion-pair, and so a total production of $\approx$ 2.3$\cdot10^3$ H$_2$SO$_4$ molecules cm$^{-3}$s$^{-1}$. This amounts to an increase of $\approx$ 7 \% in the production of H$_2$SO$_4$ molecules which is sufficient to compensate for the numerically determined $\approx$ 3-7 \% decrease in  H$_2$SO$_4$ concentration caused by the increase in number of aerosols. Recent \textit{ab inito} calculations have confirmed the first steps of such reactions in small clusters\cite{Bork2011ACP,Bork2011ACPB,Bork2012}.

So in conclusion it has been shown that an increase in ion-induced nucleation survives as the clusters grow into CCN sizes in direct contrast to the present neutral experiment and current theoretical expectations. It is proposed that an ion-mechanism exists which provides a second significant pathway for making additional H$_2$SO$_4$, as a possible explanation of the present experimental findings. Irrespective of the detailed mechanism leading to the results presented here they provide a possibly important missing piece of the puzzle as to why responses in aerosol to variations in ionization have been seen in cloud properties.

The authors thank M. Avngaard and J. Polny, for their technical assistance in relation to the experimental setup, and we thank Anne Gry Hemmersam and Morten Lykkegaard, Danish Technological Institute, Aarhus, for lending us the TSI Model 3480 Electrospray Aerosol Generator. HSV thanks Nigel Calder for stimulating discussions. Finally we thank the Carlsberg Foundation.


\end{document}